\documentclass[conference]{IEEEtran}

\usepackage{trackchanges}
\addeditor{vg}
\addeditor{dz}
\addeditor{zahra}

\ifCLASSINFOpdf
\else

\fi

\usepackage[ruled]{algorithm2e}
\usepackage[noend]{algpseudocode}
\usepackage{stmaryrd}
\usepackage{marvosym}
\usepackage{booktabs}
\usepackage{enumitem}
 \usepackage{moresize}
\usepackage{dblfloatfix}  
\usepackage{varwidth}
\usepackage{capt-of}
\usepackage{listings}
\usepackage{amsmath}
\usepackage{framed}
\usepackage{tabu}
\usepackage{float}
\usepackage{enumitem}
\usepackage{color}
\usepackage{graphicx}
\usepackage{cleveref}
\usepackage{fancybox}
\usepackage[export]{adjustbox}
\usepackage[table]{xcolor}
\definecolor{Gray}{gray}{0.9}
\definecolor{White}{RGB}{255,255,255}
\usepackage{bbold}
\usepackage{bbm}
\usepackage{mathbbol}
\usepackage{multirow}
\usepackage{fix-cm}
\SetKwInput{KwData}{Input}
\usepackage{siunitx,booktabs}
\algnewcommand\algorithmicforeach{\textbf{for each}}
\algdef{S}[FOR]{ForEach}[1]{\algorithmicforeach\ #1\ \algorithmicdo}
\usepackage{xcolor}
\def\headline#1{\hbox to \hsize{\hrulefill\quad\lower.3em\hbox{#1}\quad\hrulefill}}
\usepackage{setspace}
\usepackage{graphicx}
\usepackage{colortbl}
\usepackage{array}
\usepackage{pifont}
\usepackage {rotating}
\usepackage{mwe}
\usepackage{pbox}
\usepackage{enumitem}
\let\oldding\ding
\renewcommand{\ding}[2][1]{\scalebox{#1}{\oldding{#2}}}
\setlist[description]{leftmargin=\parindent,labelindent=\parindent}

\newcommand{\Csharp}{%
  {\settoheight{\dimen0}{C}C\kern-.05em \resizebox{!}{\dimen0}{\raisebox{\depth}{\#}}}}

 \usepackage{enumitem}
\usepackage{xspace}
\usepackage{xparse}
\DeclareDocumentCommand\newstep{o}{%
\item\IfNoValueTF{#1}{}{#1 \textendash\xspace}}

\newlist{steps}{enumerate}{1}
\setlist[steps]{label=\textit{Step \arabic*:},leftmargin=*}
 \usepackage[style=ieee, backend=bibtex, defernumbers=true]{biblatex}

\definecolor{orange}{RGB}{0,32,96}
\definecolor{g}{RGB}{50,50,50}
\definecolor{brightmaroon}{rgb}{0.76, 0.13, 0.28}
\definecolor{byzantine}{rgb}{0.74, 0.2, 0.64}
\definecolor{chromeyellow}{rgb}{1.0, 0.65, 0.0}
\definecolor{applegreen}{rgb}{0.55, 0.5, 0.0}
\definecolor{cadetgrey}{rgb}{0.57, 0.64, 0.69}

\usepackage{capt-of}

\PassOptionsToPackage{gray}{xcolor}
\usepackage{tikz}

\usepackage{tcolorbox}

\usepackage{amsmath}
\usepackage{wrapfig}
\usepackage{siunitx,booktabs,subcaption}
\makeatletter
\def\ps@IEEEtitlepagestyle{%
  \def\@oddfoot{\mycopyrightnotice}%
  \def\@evenfoot{}%
}
\def\mycopyrightnotice{%
  {\hfill \footnotesize [Preprint version]  2018 IEEE 26$^{\text{th}}$ International Requirements Engineering  Conference Workshops (REW'18) \hfill}
}

\begin{document}

\title{Loud and Interactive Paper Prototyping in Requirements Elicitation: What is it Good for?}
\author{\IEEEauthorblockN{Zahra Shakeri Hossein Abad, Sania Moazzam, Christina Lo, Tianhan Lan, Elis Frroku, Heejun Kim}
\IEEEauthorblockA{Department of Computer Science, University of Calgary, Canada\\
Email: \{zshakeri, 
sania.moazzam1, 
scclo, 
tianhan.lan, elis.frroku, 
heejun.kim\}@ucalgary.ca}}

\maketitle
\begin{abstract}

Requirements Engineering is a multidisciplinary and a human-centered process, therefore, the artifacts produced from RE are always error-prone. The most significant of these errors are missing or misunderstanding requirements. Information loss in RE could result in omitted logic in the software, which will be onerous to correct at the later stages of development. In this paper, we demonstrate and investigate how interactive and {\em Loud} Paper Prototyping (LPP) can be integrated to collect stakeholders' needs and expectations than interactive prototyping or face-to-face meetings alone. To this end, we conducted a case study of (1) 31 mobile application (App) development teams who applied either of interactive or loud prototyping and (2) 19 mobile App development teams who applied only the face-to-face meetings. From this study, we found that while using Silent Paper Prototyping (SPP) rather than No Paper Prototyping (NPP) is a more efficient technique to capture Non-Functional Requirements (NFRs), User Interface (UI) requirements, and existing requirements, LPP is more applicable to manage NFRs, UI requirements, as well as adding new requirements and removing/modifying the existing requirements. We also found that among LPP and SPP, LPP is more efficient to capture and influence Functional Requirements (FRs). 
\end{abstract}


\IEEEpeerreviewmaketitle

\section{Introduction and Motivation}
Getting constant customer feedback is key to developing great software products. It allows software developers to identify ways to improve products to satisfy current customers, as well as to avoid making undesirable or unpopular changes that could drive customers away \cite{Feedback}. Compared to traditional software, the process of requirements capturing of mobile applications (Apps) is more challenging as mobile Apps need a high-level of user interaction and require an intelligent and intuitive UI design. While there are well-known techniques for exploring and managing requirements change during a development life-cycle (e.g. as in \cite{APSEC, SERA, ernst2012case, change}), the process of eliciting, evolving, and managing mobile Apps' requirements is still a challenging task for App developers and challenges the research community to re-evaluate the existing RE practices when dealing with mobile App development.  

In this paper, we introduce and analyze a variation of the Wizard of Oz (WOz) technique, called LPP by which clients think aloud when interacting with a paper prototype--- a technique to capture users' intentions, challenges, and concerns when working with the system \cite{Loud}. The WOz technique, as stated by Malin Wik \cite{wik2015using}, is a low fidelity prototyping technique to simulate the requirements of a system and to give an impression of how the user can interact with the system when these requirements are actually implemented. The main motivation for using this technique as a means of studying the impact of LPP in this paper is that it can elicit requirements that are less likely to be gathered by other requirements elicitation techniques \cite{Abad2017}.

Moreover, to further analyze the application of LPP in mobile App requirements' elicitation and management, in this paper, we develop a comparison between this technique and several variations of silent paper prototyping such as traditional WOz, sketching, and storyboard. Also, we present a comparison between LPP and elicitation meetings alone, without paper prototyping. To achieve these goals, we defined the following main Research Questions (RQs):
\begin{description}
\item [RQ1-] How does paper prototyping help in capturing
mobile App requirements?
\item [RQ2-] Does LPP affect the type of extracted requirements during requirements elicitation?
\end{description}
To answer these RQs,  we conducted a case study on 183 participants who worked in groups of 2-4 over the last three years, broken down into (i) 31 mobile App development teams who applied either of interactive or loud prototyping techniques, and (ii) 19 mobile App development teams who employed only the face-to-face meetings for exploring and managing their Apps' requirements. From this study, we found that while using SPP compared to NPP is more efficient to capture NFRs, UI, and existing requirements, LPP is more applicable to influence NFRs and UI requirements as well as adding new requirements and removing/modifying the existing requirements. We also found that among LPP and SPP, LPP is more efficient to capture and influence FRs.

The rest of this paper is structured as follow: Section \ref{sec:RW} details related research on capturing and managing requirements of mobile Apps as well as the application of paper prototyping in requirements elicitation. Moreover, this section provides an overview of the prototyping techniques reviewed in this paper, which is followed by our research methodology (Section \ref{sec:Method}). The results of our both studies are reported and interpreted in Sections \ref{sec:Results} and \ref{sec:LL}, respectively. We listed the threats to the validity of our results in Section \ref{sec:TV} and finally conclude the paper with a brief remark on future research in Section \ref{sec:CI}.

\section{Related Work and Preliminaries}
\label{sec:RW}

\subsection{Related Work}
This section outlines related research on capturing and managing requirements using paper prototyping techniques. 

Vijayan and Raju \cite{Vijayan2010} conducted several case studies using student projects as data sets to examine the effectiveness of using the paper prototyping method for requirements' elicitation task. They showed that errors in the system are attributed mostly to poor communication between the user and the analysts. To resolve this, they proposed a new approach of using paper prototyping for requirements elicitation. In a similar study, to quicken the requirements elicitation process, Schneider \cite{schneider2007} developed a {\em fast feedback} technique, applying the By-Product Approach by using a custom-made tool that allows the interviewer to discuss use cases and simultaneously draw an interactive UI mock-up for the stakeholders. This allowed him/her to extract requirements that would have otherwise taken several sessions of several hour-long meetings, in a two-hour long session.

In a recent study, Abad et al. \cite{Abad2017} conducted a case study on 13 mobile App development teams that used WOz for early-stage requirements elicitation. Following the results of this study, the WOz technique proved specifically helpful in making the UI and usability of the app better, giving the teams a better understanding of what they had missed in regards to NFRs. Moreover, the authors performed manual and automatic data analysis on 40 similar apps on Google Play to compare the results with the WOz technique. They found that user reviews are a powerful tool to understand FRs, but they come at a cost. WOz proved more helpful and more cost effective when it came to NFRs. Likewise, Sefelin \cite{sefelin2003paper} carried out experiments to investigate the differences in usability and preference in paper-based low fidelity prototypes and computer-based prototypes. These experiments were conducted on a group of subjects using two systems, each with two prototypes: one paper-based, the other computer-based. The subjects were asked to perform tasks using a randomly selected prototype, whilst thinking aloud. Afterward, they were led to critique the systems and fill out a questionnaire regarding their preference. The results concluded that the critique was not affected by the type of prototype used. Although, the subjects did have a preference for the computer-based system as it made them feel less observed, and it did not cause unnecessary work for the facilitator.

Svanaes \cite{svanaes2004} conducted a day-long workshop inviting users to develop the functionalities of a mobile system through a process of scenario-building, role-playing, and low fidelity prototyping. This process attempted to limit the influence of facilitators and developers at a minimum. The results indicated that allowing the users to construct solutions according to their own real-life scenarios by restricting the low fidelity prototype to that of their desired mobile system (a paper-based prototype of a tablet device), and minimizing the influence of facilitators and developers (experts in the field), the users developed creative approaches to finding the solution, rather than the {\em correct} approach already accepted in the field. 
Manio's \cite{mannio2001} analysis of the types of prototypes, their functionalities in industrial, and commercial contexts found a general process for prototyping composed of a formula of steps. The process itself consists of seven phases, executed iteratively in a loop. Studies were conducted using a system analyst and session participants, with a combination of prototypes, and use cases, with the aim of capturing the software requirements. Following the results of these studies, requirements elicitation through prototyping was helpful in eliciting requirements from the customers, adding increased knowledge of the requirements, and developing a mutual understanding between the system analysts and the participants.

While the existing research provided valuable insights on the application of paper prototyping for requirements capturing and management, we could not find any study that investigated the application of loud WOz (i.e. LPP) and its comparison with other variations of paper prototyping in the context of RE. Moreover, the comprehensiveness of our exploratory and comparative study (i.e. running the study on 50 mobile app development teams) makes it different from other investigations.

\subsection{Paper Prototyping}
\label{sec:PP}
Three paper prototyping techniques were used by our teams to create a low fidelity prototype of their application. These techniques focus on the functional design of the application in the form of layout and button placement, and they omit any other visual details such as color, font, and images. Both of the teams using LPP and SPP were required to use all three paper prototyping techniques. The first two techniques, sketching, and storyboard, were performed identically in LPP and SPP whereas the third technique, WOz, was performed differently depending on which prototyping techniques the teams used. In this section, we detail the techniques used to design and develop our studies.

\begin{figure*}[ht]
\centering
\vspace{-3mm}
\begin{subfigure}{0.25\textwidth}
  \centering\includegraphics[scale=0.33]{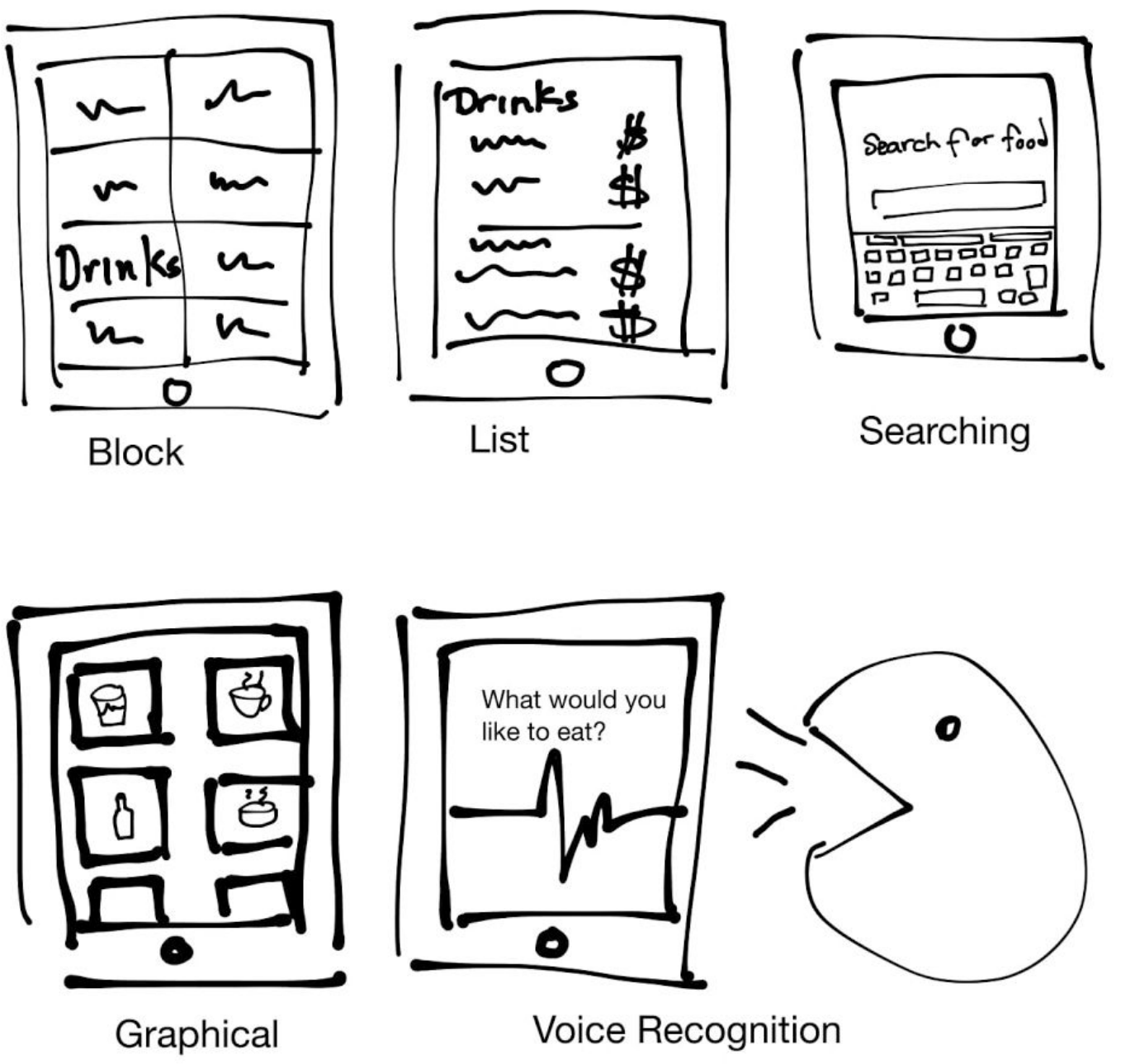}
  \caption{Sketching}\label{fig:sub1}
\end{subfigure}
\begin{subfigure}{0.25\textwidth}
  \centering\includegraphics[scale=0.5]{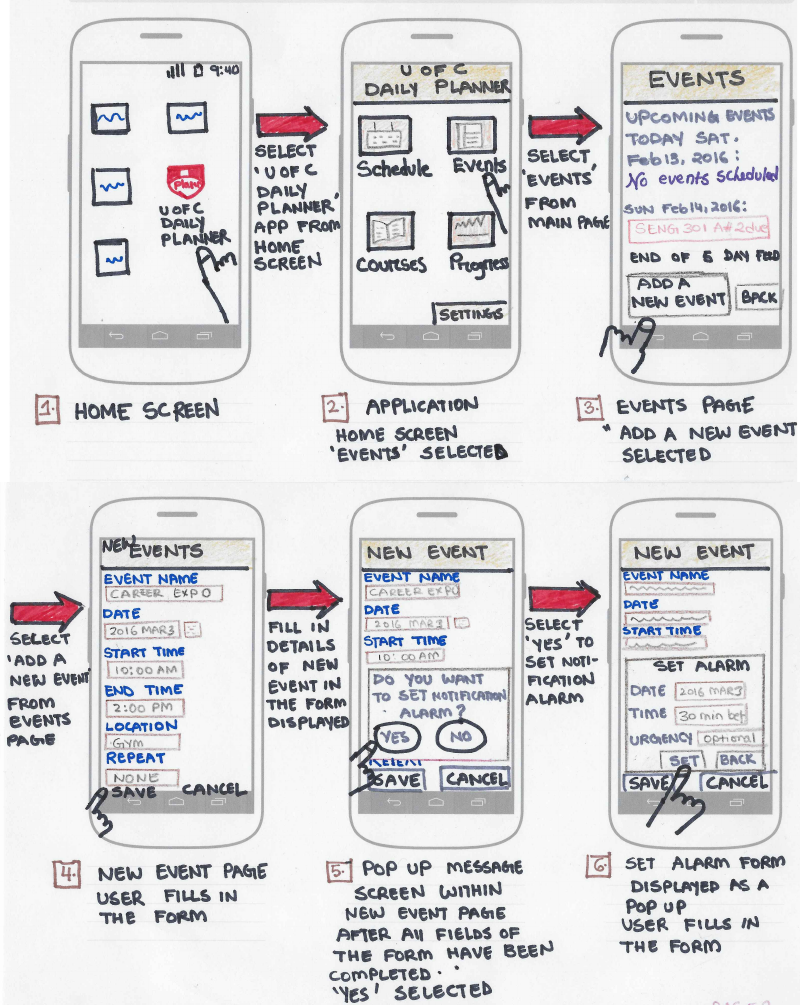}
  \caption{Storyboard}\label{fig:sub2}
\end{subfigure}
\begin{subfigure}{0.45\textwidth}
  \centering\includegraphics[scale=0.17]{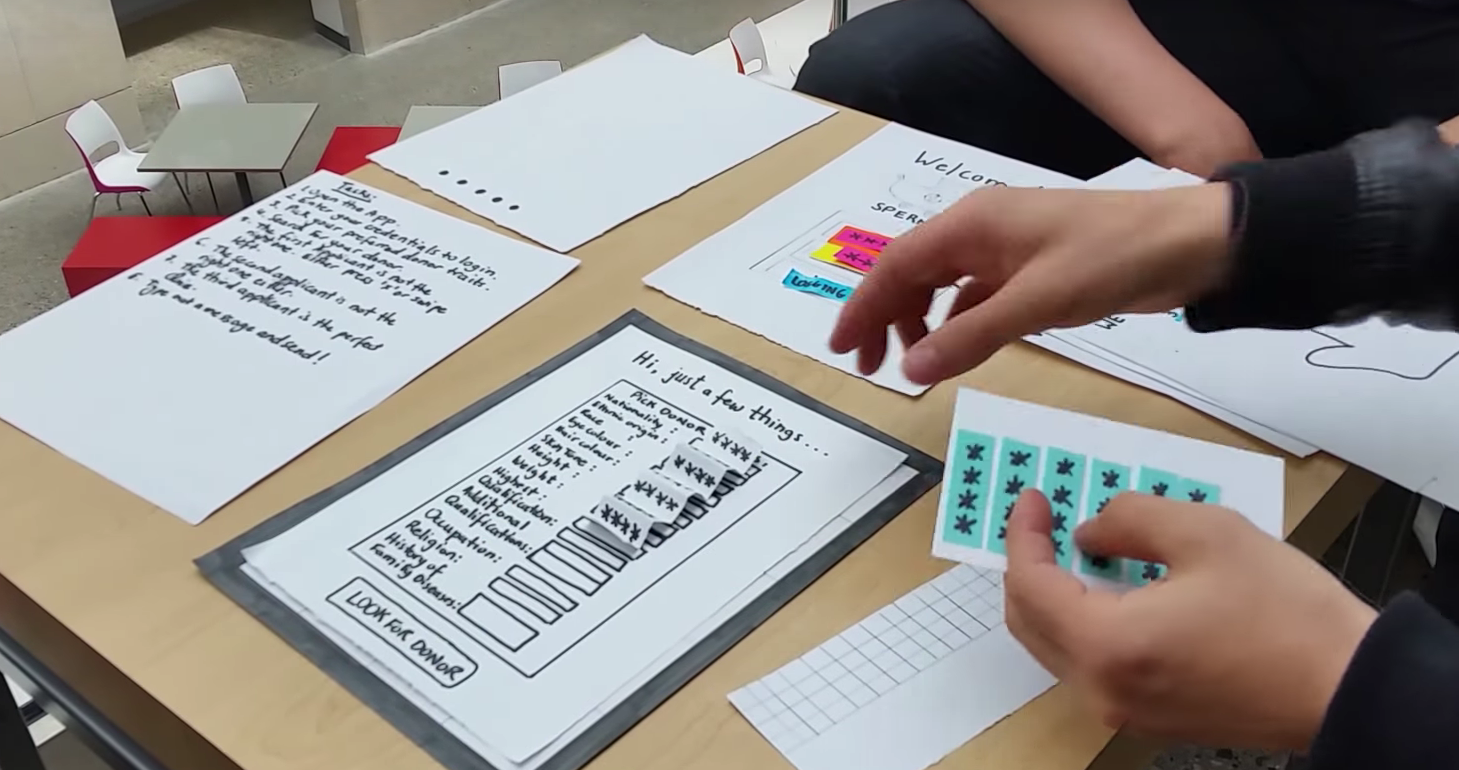}
  \caption{Loud Wizard Of OZ}\label{fig:sub2}
\end{subfigure}
\caption{The paper prototyping techniques reviewed in this paper}
\label{fig:prototypes}
\vspace{-3mm}
\end{figure*}

\subsubsection{Sketching}
The sketching technique focuses on brainstorming different ideas for the UI of the App through simple pencil and paper sketches. It is important to only use pencil and paper for this step so that it forces the team to concentrate on the functionality of each element rather than the potential visual aspects of the final product \cite{schneider2007}. In our study, teams were told to sketch five individual overview snapshots of their Apps' UI, each representing a different idea of the layout. Afterward, the teams were told to choose a single snapshot out of their five overview snapshots based on the functionality of their designs. They were then told to sketch another five snapshots to elaborate on their chosen design. The final sketches were submitted with labels under each sketch and a summary of why the final layout was chosen (see Figure \ref{fig:prototypes}a as an example of the sketching task).

\subsubsection{Storyboard}
The purpose of the storyboard is to gather information about the interactions of a software and its behavior. It is defined by Mannio \cite{mannio2001} as interactive screen displays of system behaviors that can be used for simulating man-machine behaviors. The teams are told to sketch a series of snapshots of their Apps' UI, the series representing a flowchart of what the UI would look like as a user interacts with it when performing a certain task. Each snapshot shows a current state of the UI, as well as the user interaction that leads to the next snapshot. For example, an arrow pointing to an icon would represent that the user is tapping on the icon. Similar to the sketching technique, the final sketches are labeled and a brief explanation of the steps of the task performed is included in the submission (see Figure \ref{fig:prototypes}b).



\subsubsection{Wizard of Oz}
WOz tests the usability of the design layout and detects any inaccurate user preference assumptions made during earlier design stages \cite{Dow2005} by allowing the client to test the UI through interacting with a paper prototype as a user. A working paper model of the App is made by sketching a series of snapshots of the UI onto different sheets of paper. These snapshots map out what the UI would look like as a user performs a series of tasks. The model is then animated by a team member or facilitator by moving and replacing the sheets of paper as the user, who is the client, interacts with the App. The client is asked to perform a series of tasks with the paper model without any or very little knowledge of how the App was designed. To implement LPP, in the third iteration of this study, the client was asked to think aloud when interacting with the App, so that the team is aware of the client's finger motions, the client's thoughts as they are navigating the system, as well as whether or not the client is struggling to complete the task. Likewise, in the first and second iterations, the client is only asked to interact with the App, without giving any comment during the interaction time. Moreover, a video was taken of the entire interaction between the client and the App. The video and the client’s comments were used to further analyze the approach. Figure \ref{fig:prototypes}c shows an example of LPP, which allows the team to communicate with the client better by allowing the client to see what the system would look like, and how interactions between the client and the system would be like.


\section{Methodology}
\label{sec:Method}
To achieve our study goals and to answer our RQs we designed and implemented a case study of (i) teams using LPP, (ii) teams using SPP, and (iii) teams using the face-to-face meetings alone, without interactive paper prototyping. 
\subsection{Tasks and Data Collection}
The data for implementing our case study was gathered during a three-week study in three different years (2016-2018) from 50 mobile App development teams, totaling 183 students during the beginning stages of the Android application development. The teams were provided with two weeks of instruction on RE and low fidelity prototyping. In the following two weeks, the mobile App members from each team applied their knowledge accrued during the instruction phase to the requirements collection phase. Requirements were gathered from the initial client ~\footnote{To implement this study we recruited two categories of clients, including undergraduate software engineering students and professional software developers from industry.} descriptions and meetings between the clients and the development team (which serves as a baseline for comparison with the prototyping techniques used in this paper).  Subsequently, the development teams employed WOz to confirm all requirements were collected from the client and were understood correctly by the team. WOz was carried out through the interaction between the client, the development team, and the low fidelity model prototype. Any changes or additions made to the requirements were submitted to the research team as responses.

Moreover, to organize the process of our data collection, we created three data collection forms, in which the following data points were captured for each team under study: (1) the type of the paper prototyping - loud or silent (2) the type of the influenced requirements - FR, NFR, or UI (3) the impact of the prototyping technique on the existing/potential requirements - add, remove, or modify. The collected data, using these data extraction forms, were used to address RQ2 (i.e. to run statistical tests, Section \ref{sec:SA}). Also, to reduce the data collection bias, the recorded data for each form has been reviewed by at least two authors of the paper. 

\subsection{Data Analysis and Preparation}
\label{sec:DC}
Collected data were analyzed and interpreted by applying the following methods.
\subsubsection{Statistical Analysis}
To test for the impact of LPP and the differences between this technique and SPP as well as NPP (RQ2), we use the non-parametric Kruskal-Wallis test, as we could not confirm the normality of the results of our collected data for RQ2 (using {\em Q-Q} plots \cite{QQ}). To determine the statistical significance we use the p-values and report as significant, differences at 95\% confidence interval, which we use to compare the influence of paper prototyping on requirements capturing and management in above-mentioned techniques.

\subsubsection{Open Coding}
Open coding is an iterative process of manually analyzing qualitative data \cite{opencode}, which requires distinguishing and grouping interesting and similar text together. To implement this process, and to coalesce all of the references, we used the NVivo \cite{Nvivo} tool, a qualitative analysis software package to code and analyze qualitative data files, obtained from 50 mobile App teams. The data files were distinguished into three groups: LPP, SPP, and NPP.  In our first meeting, we discussed the general topics of the files we had extracted and what possible nodes we would expect to see in our data. The question we were trying to answer was: {\em How does paper prototyping help in capturing mobile App requirements?}, defining our two main nodes to be FRs and NFRs. For the first iteration, the three authors of the paper coded the data files within these two nodes and kept adding subcategories as they analyzed the files. During the iterative process of open coding, we found {\em usability} and {\em learnability} to be integral parts of the NFR category, and {\em UI} and {\em interaction} to be very well defined in our data files, hence we added them under the usability node. Since there were many instances of `requirements modifying', in subsequent meetings it was decided to add `modify' as a node. Similarly, `add', `remove' and `improvement' were also added to our codings. After three iterations and several nodal adjustments later, we finally generated a hierarchy of concepts represented in a Treemap \cite{Treemap} visualization (Figure \ref{fig:TreeMap}).

\subsubsection{Natural Language Processing}
Natural Language Processing (NLP) is the analyzing of human language with a computer program, to derive and form understanding without human supervision. For our analysis, we used the NLP algorithm Latent Dirichlet Allocation (LDA) to analyze 31 text files (obtained in data collection, from 31 mobile App development teams who applied either of interactive or loud prototyping techniques). This algorithm analyzed the dataset by modeling the text into a number of topics (i.e. $k$). The value of $k$ was an initial value defined by us as $k$ = 3, 4, 5 (to model the dataset into 3, 4, and 5 topics). However, we later determine the $k$ value that is best suited for modeling our dataset by analyzing the coherency of the resulting topics formed by each of the three $k$ values (Section IV-A-2). A topic, as defined in the LDA approach, is a probability distribution over a vocabulary \cite{blei2010}, which required the \textsf{topicmodels} package in \textsf{R} for the correct implementation of the algorithm. Furthermore, the Gibbs sampling option was used with the LDA algorithm because it acquires higher accuracy than the variational algorithm \cite{Porteous2008}.

In order to analyze our dataset with the LDA algorithm, we first prepared our data by combining and stripping the dataset into one large group of unlabeled text. To do this, we performed the following formatting steps to the dataset:

\begin{enumerate} 
\item {\em Conversion to Lowercase:} We started our preprocessing by transforming the text to lower case to remove case-sensitivity and streamline the following stop-word removal processes.
\item {\em Removal of Non-English Vocabulary Text:} Non-English vocabulary text such as HTML formatting and tags, punctuation, and digits, were removed. This was done through using regular expression to isolate and remove HTML tags (``$<.*?>$'') and any non-alphabetical character (``[\textasciicircum a-z,A-Z ]'').
\item {\em Removal of Stop-words:} Stop-words, which are commonly used words that have no real meaning of their own such as {\em a, the, that}, were removed. This was done by using the default set of stop-words for the English language included in the package tm\_maps for R.
\item {\em Removal of Whitespace:} Any excessive whitespace such as double spaces, tabs, and newlines were removed and replaced with a single space.
\item {\em Manual Transformation:} Certain words that are commonly written in both the form of one word and two words were combined to form a single word. For example, non-functional became nonfunctional. In addition to this, words that are different but have the same meaning in the context of requirements elicitation were replaced with the same word (e.g. ``client" became ``user").
\item {\em Pre-Stem Removal of Additional Words}:
Additional words were removed from the dataset, such as words that are commonly used by students but provide no meaning to the overall topic model. This process was iterative, and we would make alterations to the list of insignificant words every time we ran the code and reviewed the results. These words tend to be component-based such as “application”, “system”, and “video”, but they would not provide insight into the requirement elicitation process if placed in a topic. 
Stop-words that were not caught by the default package were also removed during this stage. Additional word removal was manually performed by a programmer by listing the words that were insignificant and using the word removal function in the package tm\_maps for R. 
\item {\em Stemming:} Stemming is a process of removing the suffixes attached to words to reduce them to their originating word for example, “interaction”, “interacts”, and “interacting” would all become “interact”. This allowed the algorithm to analyze a word with many suffix variations as a single word. 
\item {\em Post-stem Removal of Additional Words}: Any words that should have been removed by previous code yet still appear in the dataset were manually removed at this stage.
\end{enumerate}

With the dataset processed, the LDA algorithm was implemented within the parameters of the Gibbs sampling option, the dataset, and the $k$ values. The algorithm was executed three times, once for each k value.

\section{Results}
\label{sec:Results}
This section reports the key findings of our study in two parts: the application of paper prototyping in RE (RQ1), and the comparison between LPP and other approaches such as SPP, and NPP (RQ2).
\subsection{RQ1-The Application of Paper Prototyping in Requirements Elicitation}

\begin{figure}
\centering
\vspace{-3mm}
\includegraphics[scale=0.75]{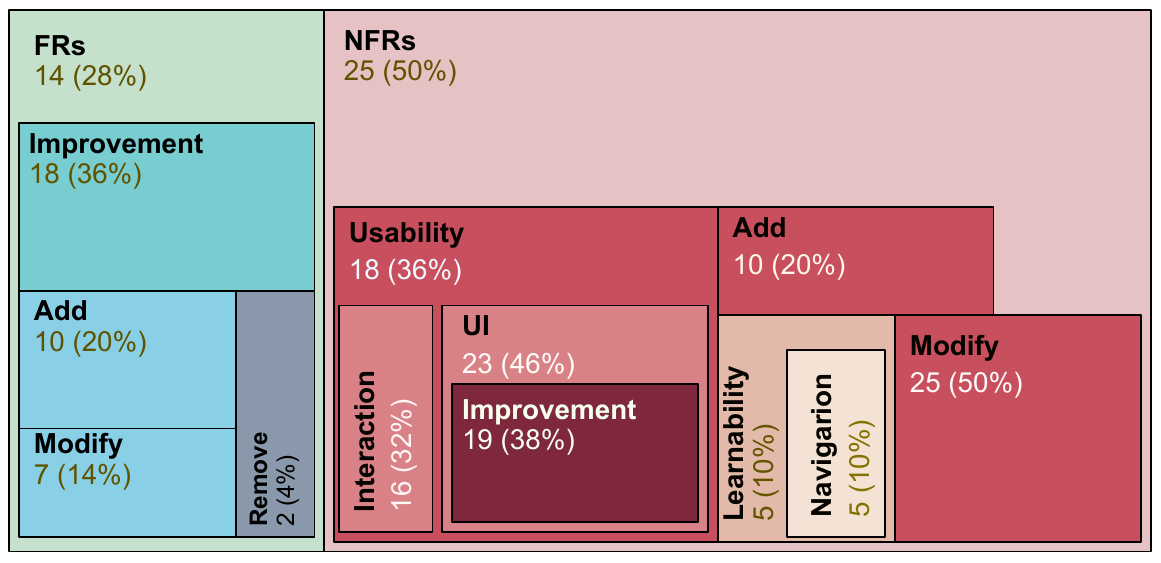}
\caption{\footnotesize The results of the manual open coding process to answer RQ1---the application of paper prototyping in requirements elicitation.}
\label{fig:TreeMap}
\vspace{-5mm}
\end{figure}


To investigate the application of paper prototyping techniques, the teams in our study were asked {\em How does paper prototyping help in capturing mobile App requirements?}. In the rest of this section, we discuss the results of applying the manual open coding and automatic data analysis on our collected dataset.

\subsubsection{Manual Open Coding}
Through the iterative process of open coding, the results were categorized based on their similarities and internal relationships. We grouped the main characteristics that show how paper prototyping helps requirements elicitation as below: 
\paragraph{Functional Requirements} 14 (28\%) teams {\em explicitly} stated that prototyping helped revise/clarify FRs of their projects. As illustrated in Figure \ref{fig:TreeMap}, we further organized our codings of FRs on their similarities, and identified four main characteristics the teams talked about when they mentioned FRs:

{\bf Improvement$_{\text{FR}}$--} 18 (36\%) teams stated that paper prototyping notably helped clarify the projects' initial requirements giving them a better idea on how to implement those features. One team of participants stated: {\em It gave us the basic tools to organize the sequence of actions that must be taken in carrying out each requirement and the relationship between them.}

{\bf Add$_{\text{FR}}$--} New requirements were added to 10 (20\%) teams' projects as a result of paper prototyping. Discussing the prototypes with their clients gave the clients an insight to the project and clarified/modified the client's vision of the project, which led some of them to add core functionalities to the project, e.g. {\em As a matter of fact, the customer suggested two more features implement: a survival game mode (where a player would solve as many math problems correctly in a row) and the ability to change the background and theme colors of the user interface} was said by one of the teams.

{\bf Modify$_{\text{FR}}$--} After reviewing a prototype, 7 (14\%) teams' clients revised the functionalities of the project as they now had a better understanding of their product, as stated by one of the teams under study: {\em[...] For example, we emphasized a constant, live stream of data, but our client explained that we should only begin displaying data after pressing the record button. Rather than averaging readings, the application should function more like a counter over some time interval}.

{\bf Remove$_{\text{FR}}$--} Paper prototyping helped remove 2 (4\%) teams' FRs which made the project more sensible and effective. One team stated: {\em After constructing our prototypes and discussing with our clients, we collectively reached the decision to only implement the happiness feature}.

\paragraph {Non-Functional Requirements} In addition to revising FRs, 25 (50\%) of the groups stated paper prototyping was incredibly helpful in capturing NFRs. Figure \ref{fig:TreeMap} shows the different aspects prototyping helped to revise/clarify as subcategories defined within NFR:

{\bf Usability--} For 18 (36\%) of the teams that participated in our study, prototyping enhanced the usability aspects of their product. A team noted: {\em This helped clarify some details in our NFRs, such as how important it is to have first-time user instructions. During the Wizard of Oz phase, our client helped us refine our non-functional requirement by suggesting a more user-friendly interface when adding tasks.} The interaction between clients and 16 (32\%) teams' low fidelity prototypes visually communicated the areas that needed improvement. One of the responses we received was: {\em Wizard of oz helped us see how a new user will interact with our system while also getting input from our client.} Furthermore, 23 (46\%) teams talked about clarifications and 19 (38\%) teams talked about improvements made to the (UI) of the product, as a result of prototyping.

{\bf Learnability and Navigation--}  For each of the learnability and navigation categories, we received 5 (10\%) responses that indicated that prototyping helped improve a first-time user's navigation of the system, and their ability to learn/understand the system's behavior. One team wrote: {\em From the sketches, we concluded that in order to fulfill the requirement that the layout of each screen will be clean and simple ensuring easy navigation for users as well as making the App fast and easy to learn.}

{\bf Modify$_{\text{NFR}}$--} Prototyping assisted 25 (50\%) teams and their clients in modifying the initial NFRs to make the product more intuitive, interactive and user-friendly. {\em The Wizard of Oz technique helped us greatly as we saw a user using our app and observed what mistakes we had made and what feature could be made better.} was recorded by one team.

{\bf Add$_{\text{NFR}}$--} 10 (20\%) teams added new NFRs after clients interacted with the low fidelity prototypes. One of the teams stated: {\em We didn’t include back buttons initially to the game, making it impossible to change difficulty in the middle of the game or to head to the main menu.} The prototyping techniques helped team participants witness what lacked in their product and helped make adjustments and improvements to their systems.
\begin{table*}
\vspace{-2mm}
\scriptsize
 \caption{RQ1- Topic Modeling Results ({\bf K} represents the number of topics)}
 \centering
  \begin{subtable}{3.8cm}
  
     \begin{tabular*}{3.8cm}{@{}l 
        @{\extracolsep{\fill}} SS
        S[table-format=1.5(0)]
        S[table-format=1.5(0)]@{}}
          \toprule
          \phantom{Var.} &  
          \multicolumn{2}{c}{\bf K=3}\\
          \cmidrule{1-3}
            {\bf Topic \#1} & {\bf Topic \#2} & {\bf Topic \#3}\\
           \cmidrule(lr{0em}){1-3}
{client}&{function}&{sketch}\\
{create}&{help}&{design}\\
{change}&{nonfunctional}&{technique}\\
{interact}&{use}&{make}\\
{like}&{interface}&{implement}\\
          \bottomrule
     \end{tabular*}%
     \caption{}
     \end{subtable}%
      \hspace*{0.75cm}%
     \begin{subtable}{5cm}
     \begin{tabular*}{5.5cm}{@{}l 
        @{\extracolsep{\fill}} SS
        S[table-format=1.5(0)]
        S[table-format=1.5(0)]@{}}
          \toprule
          \phantom{Var.} &  
          \multicolumn{3}{c}{\bf K=4}\\
          \cmidrule{1-4}
            {\bf Topic \#1} & {\bf Topic \#2} & {\bf Topic \#3} & {\bf Topic \#4}\\
           \cmidrule(lr{0em}){1-4}
           {client} & {function} & {sketch} & {nonfunctional}\\
           {client} & {help} & {design} & {screen}\\
           {change} & {technique} & {use} & {visual}\\
           {example} & {interact} & {like} & {button}\\
           {nonfunctional} & {interface} & {menu} & {log}\\
          \bottomrule
     \end{tabular*}%
     \caption{}
     \end{subtable}%
     \hspace*{.85cm}%
     \begin{subtable}{6cm}
     \begin{tabular*}{6cm}{@{}l 
        @{\extracolsep{\fill}} SS
        S[table-format=1.5(0)]
        S[table-format=1.5(0)]@{}}
          \toprule
          \phantom{Var.} &  
          \multicolumn{4}{c}{\bf K=5}\\
          \cmidrule{1-5}
            {\bf Topic \#1} & {\bf Topic \#2} & {\bf Topic \#3} & {\bf Topic \#4} & {\bf Topic \#5}\\
           \cmidrule(lr{0em}){1-5}
           {help} & {client} & {function} & {design}&{sketch}\\
           {use} & {create} & {interface} & {change}&{make}\\
           {feature} & {nonfunctional} & {technique} & {nonfunctional}&{like}\\
           {screen} & {order} & {need} & {plan}&{interact}\\
           {button} & {base} & {implement} & {detail}&{decide}\\

          \bottomrule
     \end{tabular*}%
 	\caption{}
     \end{subtable}%
\vspace{-3mm}
\end{table*}

\subsubsection{Automatic Text Analysis}
Regarding the results of our automatic data analysis, Table I displays the results of the NLP topic modeling for $k=3,4,5$ in which every topic contains five words that should form a clear underlying concept. From evaluating the words and constructing the topics, we noted that the words in Topic \#4 of $k=4$ and Topic \#1 of $k=5$ did not form a clear underlying concept. Furthermore, Topic \#2 and Topic \#4 of $k=5$ contained a similar underlying concept of adaptability. From this analysis, we determined that exploring three topic models ($k=3$) resulted in the best representation of our dataset. The three topics revealed in the topic modeling results are:
\begin{enumerate} 
\item {\em Requirements Change ---}
(client, create, change, interact, like): our first clue in determining this category is the
frequency of the words {\em change} and {\em create} in this topic followed by {\em interact}. In the context of mobile App development, seemed like comments on creating or changing the requirements of the App received during the interaction with the prototype. 
\item {\em Usability---}
(function, help, nonfunctional, use, interface): this category represents the usefulness of the proposed UI for using/interacting with FRs and NFRs of the App.
\item {\em Visual Presentation---}
(sketch, design, technique, make, implement): in essence, this category models the requirements related to the visual appearance of the Apps, as {\em sketch} and {\em design} were the most frequent terms occurred in this topic.
\end{enumerate}

\subsection{RQ2-Loud vs. Silent Paper Prototyping}
\label{sec:SA}
\begin{figure*}[!h]
\centering
\includegraphics[scale=0.5]{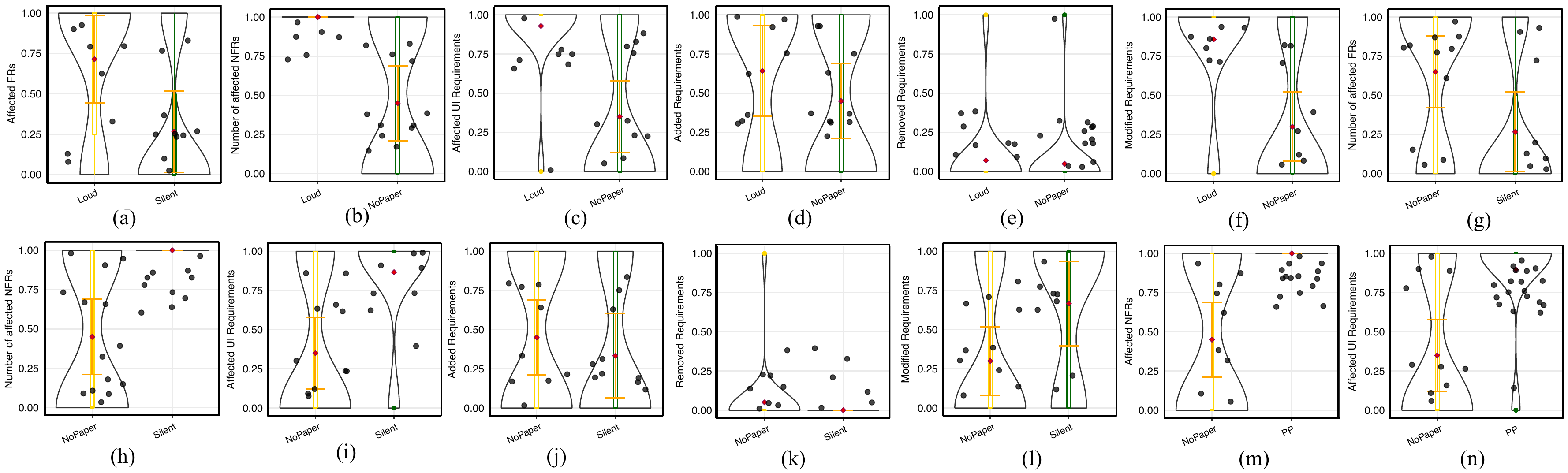}
\vspace{-5mm}
\caption{\footnotesize RQ2---95\% confidence interval of sample means for the influence of different types of paper prototyping in capturing and managing mobile Apps' requirements.}
\label{fig:Main}
\vspace{-2mm}
\end{figure*}

To answer this RQ, we posed a null hypothesis to explore the impact of {\em loud} paper prototyping on FR/NFRs, UI requirements as well as added, removed, and modified requirements of a system. We formulated this hypothesis as below:
\begin{framed}
{\em $H_0$= The loud paper prototyping technique does not impact the requirements capturing process.}
\end{framed}
To test this hypothesis, we used the dataset collected manually, using extraction forms (Section \ref{sec:DC}). Following the results of the Kruskal-Wallis tests presented in Table \ref{tab:tests}, LPP makes a significant difference in regards to influencing the FRs of the system than SPP ({\em p-value $<$0.05}). Looking at Figure \ref{fig:Main}a, for the projects under study, LPP has been more influential on managing FRs than SPP, as stated by one of our participant teams: 
{\em[...] The application of LPP (i.e. loud WOz) proved to be fruitful in exposing details in the non-functional requirements that we needed to rethink. For instance, we had not originally considered what would happen for an incomplete interval when a user presses the send log button. During the interaction of our client with the system, we determined that the log would prompt the user to fill in a task if they chose to end their working day and 15 minutes had elapsed since the last interval. 

This change allows the log to be more flexible for the user by not locking them into waiting for the interval to end. There was one non-functional requirement that became apparent from this interaction. The client realized that they wanted the user to be able to set a standard workday so that the App if forgotten at the end of the day, would automatically end the log at the specified time so that it would not run on.} 

Regarding the comparison between LPP and NPP, the results of our statistical analysis show that there is a significant difference between both the approaches in terms of their influence on NFRs, UI, as well as added, removed, and modified requirements. (Table \ref{tab:tests}). As illustrated in Figure \ref{fig:Main}b-f, in all cases there is a significant difference between LPP and NPP, LPP being more influential during the requirements capturing and management process.

Interestingly, in cases that there is a significant difference between SPP and NPP, NPP outperforms SPP in influencing FRs, adding new requirements, and removing the existing requirements (Figures \ref{fig:Main}g,j,k). Another point of interest is that SPP is more efficient to influence NFRs, UI and existing requirements of a system (Figures \ref{fig:Main}h,i,l). A possible explanation of this is that clients are more likely to have thoughts about the functionalities of their system (i.e. their motivation for soliciting its creation), as opposed to ways to judge how the system should operate.

\begin{table}
\centering
\footnotesize
\caption{\footnotesize RQ2- Comparison between various techniques of requirements elicitation. Loud Paper Prototyping (LPP), Silent Paper Prototyping (SPP), No Paper Prototyping (NPP), Paper Prototyping (PP), Functional Requirement (FR), Non-Functional Requirement (NFR), User Interface (UI).}
\label{tab:tests}
\begin{tabular}{p{2cm}cccccc} \hline
&\multicolumn{1}{c}{FR}&\multicolumn{1}{c}{NFR}&\multicolumn{1}{c}{UI}&\multicolumn{1}{c}{Add}&\multicolumn{1}{c}{Remove}&\multicolumn{1}{c}{Modify}  \\ \hline
{ \bf LPP-SPP}         &  \cellcolor{Gray}0.01 & **   &  0.6&  {0.1}&   0.3&   {0.2} \\ 
{\bf LPP-NPP}       &  0.7&  \cellcolor{Gray}{0.001}   &  \cellcolor{Gray} 0.0001& \cellcolor{Gray} 0.0001   &    \cellcolor{Gray} 0.01&    \cellcolor{Gray} 0.01 \\ 
{\bf SPP-NPP} &    \cellcolor{Gray}0.02&\cellcolor{Gray}0.001   &  \cellcolor{Gray}0.002&  \cellcolor{Gray}{0.001}   &   \cellcolor{Gray}0.0002&\cellcolor{Gray}{5e-5}\\ 
{\bf PP-NPP}                      &  0.2&\cellcolor{Gray}{0.01}   & \cellcolor{Gray}{0.001} &   0.8  &    0.7 & \cellcolor{Gray}{0.001}  \\ 
\hline 
\multicolumn{5}{l}{\em **all observations are in the same group}&\\
\end{tabular}
\vspace{-5mm}
\end{table}

\section{Lessons Learned}
\label{sec:LL}

From the NLP results, we found it interesting that the topic {\em Requirements Change} contains the word {\em client}. This suggests that flexibility from LPP is not experienced solely by the developing team but also by the clients. Since the clients are suggesting changes based on the interaction with the UI, there is a higher chance that these changes are related to the NFRs of the systems. The open coding results showed that 50\% of the teams talked about NFRs while only 28\% mentioned FRs in their data. Thus, the NLP and open coding results both suggest that LPP helps modify NFRs over FRs.
 
Another interesting result found through open coding was that out of the teams that used LPP, twenty-five teams modified NFRs while only seven modified FRs; two teams removed FRs, and no teams removed NFRs. This can be attributed to the clients' interactions with the systems using LPP resulting in the clients' realizations of how different non-functional aspects could be improved by the teams. The teams then make modifications to the NFRs of their systems whereas the overall functionalities stayed the same. Contrasting to this, with SPP for example, sketches helped clarify and remove FRs as the teams prioritized the requirements, removed what was not necessary, and understood what was needed - as they gained a better idea of their systems. From this analysis we determined that by using LPP, clients tend to suggest more changes to NFRs than FRs, leading teams to more frequently change their NFRs rather than removing them; and more frequently remove FRs rather than changing them.

However, the topic {\em visual presentation} from the NLP results and the words within the topic suggest that by using LPP, teams found it easier to design and implement their systems. The topic {\em usability} also supports this as it contains the words {\em interface}, and both {\em functional} and {\em non-functional} suggesting that designing an interface through LPP helped teams with their FRs and NFRs alike. From these results, we determined that even though clients' interactions with the systems through using LPP mainly results in feedback for NFRs, the act of making prototypes using LPP helps teams clarify and implement the FRs of their systems.
 
While most of our teams found interactive prototyping (SPP and LPP) more useful for communicating the requirements with their clients, there were teams who found the {\em storyboard} technique very useful for revising their existing requirements: {\em We found that the storyboard also helped us tweak our non-functional requirements more easily because we had to trace through our interactions with the system. We were able to recognize what was good and what was not. A prominent example of this came about when we were trying to decide how the user would add, edit, and remove tasks. We had originally considered just having a separate screen that had buttons to modify the task elements. After tracing through, we noticed that they felt rather dated and elected for a more updated approach that involved either clicking the task to edit, pressing a symbol (plus sign) to add and holding down the task to show a button to delete it.}

Among the techniques reviewed in this paper, the majority of our teams found LPP the most useful approach for managing mobile Apps' requirements. This was most likely due to the ability of this approach to interacting with the proposed system directly (enhanced with clients' feedback when thinking aloud during the interaction) as opposed to just looking at a sketch or talking about the existing and potential requirements of the system.


\section{Threats to Validity}
\label{sec:TV}
We followed and adapted the case study guidelines in software engineering provided by Kitchenham et al. \cite{Empir} and Easterbrook et al. \cite{Empir2} to mitigate the limitations of case study and empirical research. However, there are a number of threats to the validity of the results presented above.

In answering RQ1 concerning the application of paper prototyping in requirements capturing and management, the methodologies of qualitative analysis as provided for open coding and automatic data analysis were adhered to with as much rigor as the circumstances of the study would allow. For example, to mitigate the effects of potential confounding factors (e.g. the implementation of each of prototyping techniques, and developer experience), we recruited undergraduate students as subjects of our study and provided them with video instruction and well-illustrated field
guides about the paper prototyping techniques under study. However, the context in which the development teams were working poses a threat to the results of our study. The development teams who participated in our study were composed of {\em software engineering} students without industry practitioners in most cases. Moreover, this was the first time these students experienced requirements capturing and requirement engineering, in general. To mitigate this risk, all participants were provided with two weeks of instruction on RE processes and techniques, and low fidelity prototyping.

Moreover, the sample size poses a potential threat (183 participants broken down into 50 teams). While the number of participants in this study is within the parameters generally accepted in qualitative analysis \cite{patton1990qualitative}, it is not \textit{clear} that these numbers will have resulted in participant saturation; the point at which the inclusion of additional participants increases the size of the data but produces no significant changes in the results.
Finally, a recent study by Abad et al. \cite{Zahra2} indicates that LDA performs poorly when attempting to extract topics from short texts. It along with the limitations of our manual transformation task during the implementation of the LDA approach follow that the validity of the results of RQ1 might be threatened. However, since the themes were narrowed down to responses by the teams, there is not much concern over the details of each class as this limitation does not significantly impact our results. Furthermore, 50 app responses were manually analyzed to validate the results of the automatic NLP (i.e. LDA approach).

\section{Conclusion and Implications}

In researching how prototyping helps requirements elicitation and if loud and interactive paper prototyping affects the type of requirements gathered, in this paper, we used statistical analysis, open coding, and NLP to analyze data from 50 mobile App development teams with the total of 183 participants. This study showed three main results: (1) teams that used LPP clarified/modified their requirements more than the SPP teams, (2) the NLP and open coding results both suggest that LPP helps modify NFRs over FRs, and (3) Among the techniques reviewed in this paper, the majority of our teams found LPP the {\em most} useful approach for managing mobile Apps’ requirements. However, there are still some reasons to use SPP, such as:
\begin{itemize}
\item When you are measuring learnability time, thinking aloud slows interactions significantly.
\item When there is a risk that clients/users will not speak aloud until they have valid thoughts about the product. 
\item When facilitators of the study are not experienced in the context and they can easily change user behavior during their interaction with the product.
\end{itemize}

In the future, we will aim to study if LPP and SPP result in better end-products than NPP. This can be done by deploying two case studies on the data; documenting the process of product development; using questionnaires, and then performing statistical analysis on the data collected from the deployed Apps. Moreover, we aim to design and develop a tool to capture clients' thought during the interaction with the system, using the {\em loud thinking} approach and incorporate this information into the requirements specification artifact.

\label{sec:CI}

\printbibliography

\end{document}